\documentclass[aps,amssymb,amsmath,prl,twocolumn,showpacs,superscriptaddress]{revtex4}

\usepackage[latin1]{inputenc}
\usepackage{graphicx}
\usepackage{bm}

\begin{document}

\title{Collective dynamics of molecular motors pulling on fluid membranes}

\author{O. Camp\`as}
\affiliation{Institut
Curie, UMR CNRS 168, 26 rue d'Ulm 75248 Paris Cedex 05
France.}
\affiliation{Departament d'ECM, Universitat de Barcelona, Avinguda Diagonal
  647, E-08028 Barcelona, Spain.}

\author{Y. Kafri}
\affiliation{Institut
Curie, UMR CNRS 168, 26 rue d'Ulm 75248 Paris Cedex 05
France.}
\affiliation{Physics Department, Technion, Haifa 32000, Israel.}

\author{K. B. Zeldovich}
\affiliation{Department of Chemistry and Chemical Biology, Harvard University, 12 Oxford St.,
  Cambridge, MA 02138 USA.}

\author{J. Casademunt}
\affiliation{Departament d'ECM, Universitat de Barcelona, Avinguda Diagonal
  647, E-08028 Barcelona, Spain.}

\author{J.-F. Joanny}
\affiliation{Institut
Curie, UMR CNRS 168, 26 rue d'Ulm 75248 Paris Cedex 05
France.}


\begin{abstract}
The collective dynamics of $N$ weakly coupled processive molecular
motors are considered theoretically. We show, using a discrete
lattice model, that the velocity-force curves strongly depend on
the effective dynamic interactions between motors and differ
significantly from a simple mean field prediction. They become
essentially independent of $N$ if it is large enough. For strongly
biased motors such as kinesin this occurs if  $N\gtrsim 5$. The
study of a two-state model shows that the existence of internal
states can induce effective interactions.
\end{abstract}

\date{\today}

\maketitle

The collective behavior of molecular motors plays a crucial role in
many biological phenomena ranging from intracellular  and
intra-flagellar transport to axonal transport~\cite{Alberts,Howard}.
Molecular motors are often classified according to their
processivity~\cite{Leibler}. Processive motors rarely unbind from
the track on which they are moving; they perform best when working
in small groups and are therefore referred to as ``porters''.
Non-processive motors unbind from the track frequently, they work
best in large groups and are referred to as ``rowers''. Examples of
``porters'' are kinesin motors which move along microtubules, while classical
myosin motors which move along actin filaments are examples of
``rowers''~\cite{Leibler}.

The classification of motors into ``porters'' and ``rowers'' is
based on their behavior when connected to a rigid or
elastic cargo. The strong coupling between processive motors leads
to an effective friction which results from motors which cannot
move because other motors are bound to the track~\cite{Leibler,Vale}. A
strong coupling between the motors indeed exists for a microtubule
pushed by kinesin motors that are bound to a surface~\cite{Howard}. It is also
important for describing myosin motors acting in skeletal muscles.
The abundance of such systems has inspired several theoretical
studies of the collective behavior of strongly coupled motors~\cite{Jacques,Frey}.

In many cases, however, this  description in terms of rowers and
porters is not adequate since the coupling between the motors is
negligible. An important class of systems where this happens is
when motors, such as kinesin, move along microtubules and carry a
load which is a lipid membrane, an ubiquitous situation in living
cells. This occurs, for example, when kinesins or dyneins carry a
vesicle along a microtubule~\cite{Howard}. Recent experiments have
also shown that kinesin motors moving along a microtubule act
collectively to pull membrane tubes from a vesicle~\cite{Leduc}.

\begin{figure}
\centering
\includegraphics[scale=1.0]{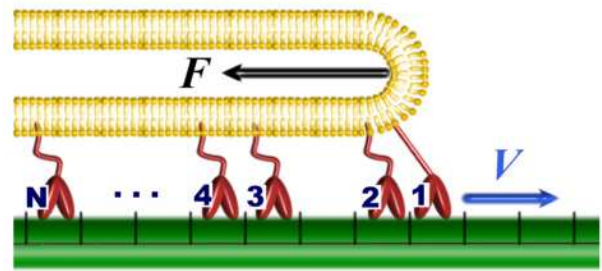}
\caption{\label{sketch} Sketch of the system. $N$ motors are pulling a
membrane tube from a vesicle. The force $F$ acts
only on the leading motor (labeled $1$). The remaining motors move
in the absence of any applied force. At long time scales, all motors move
at the same mean velocity $V$.}
\end{figure}

In this Letter, we study theoretically the collective behavior of
$N$ processive motors pulling a tube out of a membrane and acting
against the force needed to extract it~\cite{derenyi}, (Fig.~\ref{sketch}). A
fluid membrane can only exert a force on the motors
at the leading edge of the tube where the normal to the surface has a
component in the direction of motor motion.
For simplicity we assume here that all the force is transmitted to
the leading motor (Fig.~\ref{sketch}). Our treatment is  a
reasonable approximation for kinesin motors carrying a
vesicle subject to friction forces from the
cytoskeleton. It could also be relevant for possible single molecule
experiments where a bead is exerting a force on a single motor
moving in front of several other motors.

We consider the collective behavior of the motors as a function of
the applied force, $F$, the number of motors, $N$, and the
effective interactions between the motors. An effective
interaction is defined as the combined effect of the microscopic
details of the system on the transition rates in a coarse-grained
description (see discussion below). It is shown that the
velocity-force curve $V_N(F)$ strongly depends on the interactions
between motors and is different from that of a simple mean-field
treatment in which independent motors share the force $V_1(F/N)$.
Moreover beyond a certain number of motors, the velocity-force
curves are all indistinguishable for practical purposes.  While
the interactions do not play any role in the absence of external
force, their effect becomes clearly visible as $F$ increases. The
possibility of extracting the nature of the effective interactions
between motors from experiments is discussed. Finally we explore
how the effective interactions in the coarse-grained description
arise from a more microscopic two-state model.

\begin{figure}
\centering
\includegraphics[scale=0.85]{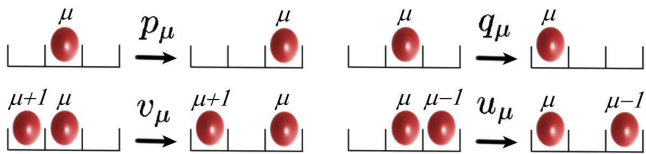}
\caption{\label{rates} Possible motor transitions and associated
  rates. The boxes and balls represent lattice sites and motors
  respectively.}
\end{figure}

In a coarse-grained description of the system we first
model the motors as interacting biased random-walkers moving
along a one-dimensional lattice. We assume that the  motors are
fully processive and never unbind from the filament (microtubule
or actin) that acts as a track. The lattice constant $\ell$
 is the periodicity of the filament. Each site can only
be occupied by one motor which can move to a neighboring
site if empty. We label the motors with an
index $\mu=1 \ldots N$, with $1$ labeling the motor on which the
force is exerted (Fig.~\ref{sketch}). The dynamics of the motors is
specified by the hopping rates defined in Fig.~\ref{rates} where
the boxes
 represent sites on the lattice and a ball with index
$\mu$ indicates that the site is occupied by motor $\mu$.

The model is a generalization of the disordered exclusion model
introduced in~\cite{Evans96} which includes modifications of the
rates due to nearest-neighbor interactions between the motors. The
hopping rates are chosen as follows: $p_\mu=p$, $v_\mu=v$,
$q_\mu=q$ and $u_\mu=u$ for $\mu \geq 2$; the rates of the leading
motor ($\mu=1$) depend on the external force~\cite{jaume}. Using
Kramers rate theory~\cite{kramers}, we assume for simplicity an
Arrhenius dependence on the force so that $p_1=pe^{-f \delta}$,
$q_1=qe^{f (1-\delta)}$ and $v_1=ve^{-f \delta}$, where $f$ is the
force in units of $k_B T/\ell$ ($k_B$ being the Boltzmann constant
and $T$ the temperature). The dimensionless parameter $0<\delta<1$
characterizes the position of the energy barrier between the two
neighboring lattice sites. Attractive effective interactions
correspond to reduced hopping rates ($v<p$, $u<q$ ) and
 repulsive effective interactions to increased
hopping rates ($v>p$, $u>q$). We refer to the case $v=p$ and $u=q$ as neutral.

It is instructive to first consider a system with only two motors.
This case may be solved exactly in the long-time limit. For any
finite force the probability of finding the motors $k$ sites apart
decays as $\left[ (p_1+q)/(p+q_1) \right]^k$. The average number
of sites between the two motors is therefore finite and decreases
with the force. Since the motors cannot overtake each other, their
velocities are equal and read:
\begin{equation}
V_2 = \frac{v_1 (p-q) + u (p_1-q_1)}{(v_1+u)+(p-q)-(p_1-q_1)} \;.
\label{steadyvel}
\end{equation}
For comparison,
the velocity of a single motor within this model is $V_1=p_1-q_1$.
The maximum force that the motors can exert (stall
force) is the force for which the velocity vanishes. The stall force of a single
motor is given by $f_s(1)=\ln (p/q)$,
while using Eq. (\ref{steadyvel}) the stall force of two motors is
\begin{equation}
f_s(2) =  \ln \left( \frac{pv}{qu}+\frac{p}{q}- \frac{v}{u}
\right) \; . \label{stall2}
\end{equation}
The stall force is not necessarily twice the stall force of a
single motor. It is a function of the rates ratio $v/u$, which
depends on the interactions between the motors, and can be either
larger or smaller than $2f_s(1)$ depending on whether $v/u>p/q$ or
$v/u<p/q$ respectively.

\begin{figure}
\centering
\includegraphics[scale=0.35]{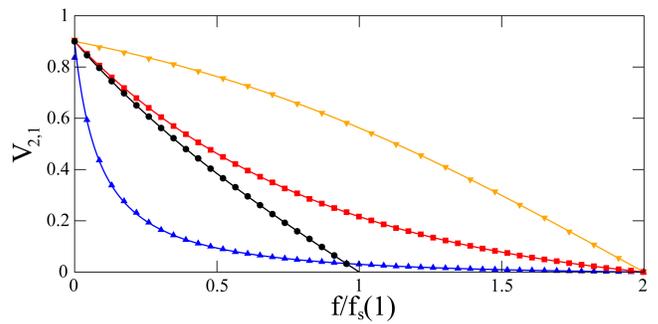}
\caption{\label{2mot_inter} Velocity-force curves of $2$ motors in the limits of
attractive, repulsive and neutral interactions. Analytical solutions of
  Eq. (\ref{steadyvel}) (solid lines) and Monte Carlo simulations (symbols).
For all cases $p=1.0$,
  $q=0.1$ and $\delta=0.5$. The rates $v$ and
$u$ are: $v=0.1,1.0,10.0$ and $u=0.01,0.1,1.0$ for attractive (triangle up),
  neutral (square) and repulsive (triangle down) interactions
  respectively. The velocity-force curve for a single motor is also plotted for
  comparison (circle). The ratio $v/u=p/q$ so that $f_s(2)=2f_s(1)$. All rates
  are in units of $p$.}
\end{figure}

The velocity $V_2$ is plotted for various values of $v$ and $u$ in
Fig.~\ref{2mot_inter} where, for clarity, we set $v/u=p/q$. The velocity $V_1$
of a single motor is also shown in
the figure. The general shape of the velocity-force curve is
highly sensitive to the interactions. For strong enough attractive
interactions the velocity of two motors is smaller than that of a
single motor up to a certain value of the force. At an interaction
dependent point the two curves cross and the two motors become
faster than a single motor for large enough forces. An
experimental signature of this type is a clear demonstration
of attractive interactions between the motors.

We now turn to the general case with $N$ motors.
 Using the result
of~\cite{Evans96}, an exact expression of the velocity can be
obtained in the neutral case where $v=p$ and $u=q$ on a ring
geometry. In the limit where the number of vacancies in front of
the first motor ($\mu=1$) is infinite, the periodic boundary
conditions do not influence the results. Building on the results
of Ref.~\cite{Evans96} one finds
\begin{equation}
V_N=p\frac{\left[1-e^f
    (q/p)^N\right]\left[1-q/p\right]}{e^{f\delta}\left[1-q/p\right]+e^f\left[
    q/p-(q/p)^N\right]} \;. \label{veln}
\end{equation}
In the neutral case, for any number of motors $f_s(N)=Nf_s(1)$.
For large $N$, the slope and therefore the velocity near stall
force, decrease exponentially with the number of motors as
$(q/p)^N$. Even
in the neutral case, the velocity-force curve is different from
the naive mean-field prediction $V_N(F)=V_1(F/N)$. In the absence
of force, the velocity is independent of the number of motors.
The slope of the velocity-force curve for vanishing forces is negative and
converges exponentially fast with $N$ to
$-(1-q/p)\left[q+(p-q)\delta\right]$. The larger the number of
motors the smaller the absolute value of the slope. In particular,
these results imply that for any $N \gg -1/\ln(q/p)$ the velocity
force curves are indistinguishable for any practical purpose.

Close to stall force, the velocity and the stall
force can be obtained in the presence of interactions in the limit
where $p \gg q$, $p \gg u$ and $v \gg u$: the motors then form a
compact cluster and the movement in either direction occurs by
propagation of a vacancy from one end to the other. This argument
leads to
\begin{equation}
\label{VN} V_N=v_1 v/q_1 - u(u/p) (u/v)^{N-3} .
\end{equation}
One can check that this is in agreement with the general result for two motors
and with the result for $N$ motors in the neutral case.
The normalized stall force for $N$ motors in this limit is then
\begin{eqnarray}
\label{stallforce} f_s (N)/(Nf_s (1)) &=& \ln (v/u)/ \ln (p/q)
\nonumber \\&-&\left( \ln (v/u)/ \ln (p/q)-1 \right)/N
\end{eqnarray}
As for two motors, the stall force only depends on
the rate ratios $v/u$ and $p/q$. If $v/u > p/q$ the normalized stall force per motor
increases with the number of motors and saturates at a value larger
than $1$ for many motors. It has the opposite behavior when
$v/u<p/q$.

\begin{figure}
\centering
\includegraphics[scale=0.35]{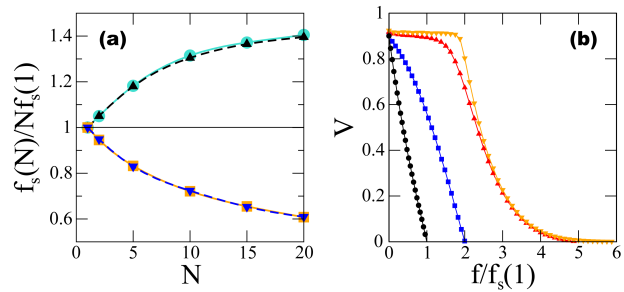}
\caption{\label{stall_n} (a) Stall force as a function of $N$ for various types of
interactions. Both for attractive ($v=0.7$, $u=0.5$; circles) and
repulsive ($v=1.54$, $q=1.1$; triangles up) interactions, the value of
the stall force, $f_s(N)$, is the same and larger
than $Nf_s(1)$ as $v/u=1.4>p/q$. When
$v/u=1.1<p/q$, $f_s(N)<Nf_s(1)$ and it has also the same value for
both attractive ($v=0.55$, $u=0.5$; squares) and repulsive
($v=1.21$, $u=1.1$; triangles down) interactions.
(b) Velocity-force curves in the case of repulsive interactions between highly
biased motors ($p=1, q=0.1, v=10, u =1$ for $1$ (circles), $2$ (squares),
$5$ (triangles up), $10$ (triangles down) motors;
$\delta =0.5$. All rates are in units of $p$. }
\end{figure}

We have also performed continuous time Monte Carlo simulations
(see e.g.~\cite{Newman}) to test the effect of interactions
between motors. Similarly to the case of two motors, when $v/u
\neq p/q$ the stall force is only a function of the ratio $v/u$
(Fig.~\ref{stall_n}a). When $v/u = p/q$ the stall force always
satisfies $f_s(N)=Nf_s(1)$ as expected. This statement can be made
rigorous by showing that with $v/u = p/q$ at the stall force
detailed balance holds~\cite{US}. The general shape of the
velocity-force curve reveals that as in the case of two motors,
when the interactions are strongly attractive, there is a
cross-over from a low force regime where the velocity is lower
than that of a single motor to a regime where it is higher. An
important result of the simulations is that, similarly to the
neutral case, for any given type of interaction the velocity-force
curves are all nearly identical above a certain number of motors.
We stress that the comparison between the velocities of many and
one motor could serve as an experimental test to sort out
attractive or repulsive interactions. In the presence of repulsive
interactions, the velocity $V_N$ of $N$ motors is always larger
than the velocity of $1$ motor and the velocity force curves all
collapse on a single curve if $N\gtrsim 7$, ($p/q=10$)
(Fig.~\ref{stall_n}b). In the presence of strong enough attractive interactions
the velocity is always smaller than that of a single motor for
small forces but becomes larger at larger forces. The
velocity-force curves collapse if $N\gtrsim 5$, ($p/q=10$).
Experimentally one should expect a velocity-force curve
independent of the number of motors if a few motors act
collectively. The observed stall force could also be much smaller
the predicted theoretical one since for many motors, the velocity
reaches negligible values way below stall force.

\begin{figure}
\centering
\includegraphics[scale=0.35]{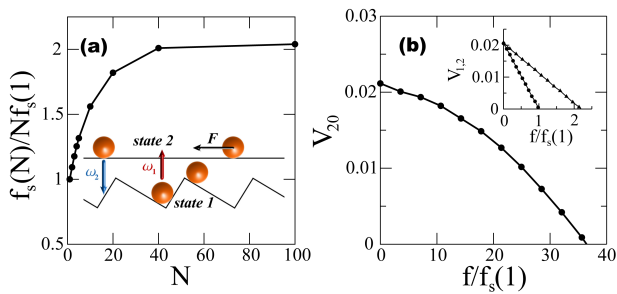}
\caption{\label{ratchet} Simulation results for motors with two internal
states (two-state model) and only excluded volume interactions. (a) Stall force as a
function of the number of motors. The value of the stall force is
larger than $N f_s(1)$. (b) Velocity-force curve for $20$ motors.
The inset in
(a) sketches the two-state model, and the one in (b) shows the velocity-force
relation for $1$ (circles) and $2$ (triangles up) motors.
}
\end{figure}

In the above discussion the nature of the effective interaction
(neutral, attractive or repulsive) between the motors was assumed.
We now argue that for motors with several internal states, one
expects generically non-neutral interactions on long times and
large length-scales. We use the example of a two-state
model~\cite{Jacques02} (Fig.~\ref{ratchet}a, inset). In the strongly bound state ($1$) the motor
feels the sawtooth potential, $W_1(x)$, with a period $\ell$, an
amplitude $5 k_B T$ and a short segment of the sawtooth of length
$a=0.2 \ell$. In the weakly bound state ($2$) the potential
$W_2(x)$ is constant. The motors change from state $1$ to state
$2$ and vice-versa with local excitation rates $\omega_{1}(x)$ and
$\omega_{2}(x)$ respectively. The transition rates, in arbitrary
units, are given by $\omega_{1}(x)=\frac{\Omega}{\alpha\sqrt{\pi}}
\exp[{-(x \mod \ell)^2/\alpha^2}]$ with $\Omega=2$,
$\alpha=0.05\ell \ll a \ll \ell$ and $\omega_{2}(x)=0.2$. We
assume only hard core interactions between the motors. These
are captured by a repulsive potential $U(y)$ which depends on the
distance, $y$, between the motors. The potential chosen is the
shifted repulsive part of a Lennard-Jones potential vanishing at
$y> 2^{1/6} \sigma =1.68 \ell$, with an amplitude
$\varepsilon=0.05 k_B T$.
 The interaction range is $\sigma=1.5\ell$.
We have verified that our results remain qualitatively the same
upon changing  the details of the potential and the values of the
parameters.

We numerically simulate the model using Langevin dynamics for the
motors. The equations for motor $\mu$ in state $s_\mu$ read
\begin{equation}
\xi \frac{dx_\mu}{dt} = - \frac{dW_{s_\mu}(x_\mu)}{dx} -
\frac{d}{dx_\mu} \sum_{\kappa \neq \mu} U(x_\mu-x_\kappa) + F
\delta_{\mu,1} +\eta, \label{LangevinEq1}
\end{equation}
where $F$ is the external opposing force, and $\xi=50$ is the
dimensionless friction coefficient of the motor. The random force
is described by the noise term $\eta=r \xi \sqrt{\frac{6 k_BT}{\xi
dt}}$, where $r$ is a random number taken from a uniform
distribution from -1 to 1. These equations are coupled to standard Monte Carlo
steps for the transitions between the bound states $1$ and $2$.
Initially, the $N$ motors are placed randomly at distances
exceeding $2^{1/6}\sigma$, so that the interaction energy $U(y)$
vanishes. Throughout the simulation, we follow the position of the
first motor and determine its velocity at long times.

The velocity-force curve obtained from the simulations for $20$
motors is plotted in Fig.~\ref{ratchet}b. Since the parameters
were chosen so that the stall force of one motor is small, the
velocity-force relation is nearly linear for a small number of
motors (Fig.~\ref{ratchet}b, inset). Increasing the number of
motors reveals the non-linearities. The comparison between the
general shape of the curve for $20$ motors (Fig.~\ref{ratchet}b)
with the ones obtained from the coarse-grained model, suggests
that the existence of two internal states for the motors leads to
effective repulsive interactions. The stall force is
plotted in Fig.~\ref{ratchet}a as a function of the number of
motors. For a given $N$, it is larger than $Nf_s(1)$, indicating
that the effective rates for forward and backward movement in the
equivalent coarse-grained model are such that $v/u>p/q$. As the
number of motors is increased this effect becomes more important
and saturates for large $N$. Overall, these results are consistent
with those obtained for weakly biased random-walkers with repulsive
interactions.

We now give a qualitative discussion of
the two-state model for two motors. A simple mean-field approach
would assume that the effect of the second motor is to contribute
with a mean force, which is that produced by the single motor at
the actual velocity of the two motors. This yields
$V_2(F)=V_1(F/2)$. In this approximation, the two motors equally
share the force. This result holds only for sufficiently smooth long range
interaction potentials.

Going beyond mean field, we have identified two configurations
that contribute most to the increase of both the stall force and
the velocity. If the leading motor carrying the force is in the
unbound state and the trailing motor in the bound state, the force
exerted by the filament potential on the trailing motor pushes the
leading motor forward. This is best seen in the case where the
diffusion constant of the motors is very small and $f>0$: the
velocity of a single motor is small, as is evident from the
ratchet potentials, but the velocity of the two motors remains
finite as the trailing motor pushes the leading motor over the
potential barrier. This effect is very sensitive to the mismatch
between $\sigma$ and $\ell$. On the other hand, if the two motors
are in the unbound state the entropic repulsion between the two
motors biases forward the motion of the leading motor. Using these
ideas one can explicitly check in the limit where $\sigma$ is
close to a multiple of $\ell$, that
 $f_s(2)>2f_s(1)$, the difference increasing
with the diffusion coefficient of the motors~\cite{US}.

In conclusion, we have shown using various models of molecular motors
that the collective behavior of a cluster of weakly coupled motors
depends on their dynamic interactions and is very different from both
the mean field prediction and from the behavior of strongly
coupled motors.

We acknowledge stimulating discussions with M. R. Evans and the
financial support of the European Commission (HPRN-CT-2002-00312),
the Spanish M.E.C. (O.C.), MCyT (Spain), project
BQU2003-05042-C02-02, and the H.F.S.P. (Y.K. and K.B.Z.).


\begin{thebibliography}{99999}
\bibitem{jaume} We assume here for simplicity that the N-motor problem can be
  described with the same rates as the two motor case: we ignore higher order
  correlations on the rates.
\end{thebibliography}
\end{document}